\def\order#1{{\cal O}\left(#1\right)}
\def\ba{\begin{eqnarray}}
\def\ea{\end{eqnarray}}
\def\vmu{\mbox{\boldmath$\mu$}}
\def\boldP{\mbox{\boldmath$P$}}
\def\boldp{\mbox{\boldmath$p$}}
\def\boldE{\mbox{\boldmath$E$}}
\def\boldB{\mbox{\boldmath$B$}}
\def\vsig{\mbox{\boldmath$\sigma$}}

\documentstyle[preprint,tighten,aps,psfig]{revtex}
\newcommand{\be}{\begin{equation}}
\newcommand{\ee}{\end{equation}}

\def\vec#1{{\mbox{\boldmath$#1$}}}

\newcommand{\p}{\mbox{$\vec{p}$}}

\newcommand{\A}{\mbox{$\vec{A}$}}

\newcommand{\PP}{\mbox{$\vec{P}$}}

\preprint{BNL-HET-00/24, SLAC-PUB-8522, hep-ph/0007217}

\begin{document}

\title{Anomalous magnetic moment of a bound electron}

\author{Andrzej Czarnecki\thanks{
e-mail:  czar@phys.ualberta.ca}}
\address{
Department of Physics, University of Alberta\\
Edmonton, AB\ \  T6G 2J1, Canada\\
and\\
Physics Department, Brookhaven National Laboratory,\\
Upton, NY 11973}

\author{Kirill Melnikov\thanks{e-mail: melnikov@slac.stanford.edu}}
\address{Stanford Linear Accelerator Center\\
Stanford University, Stanford, CA 94309}

\author{ Alexander Yelkhovsky\thanks{
e-mail:  yelkhovsky@inp.nsk.su }
}
\address{ Budker Institute for Nuclear Physics,         
\\
Novosibirsk, 630090, Russia}
\maketitle

\begin{abstract}
We study binding corrections to the gyromagnetic factor $g_{\rm e}$ of
an electron in hydrogen-like ions.  We argue that the leading order
binding effects in radiative corrections $\Delta g_{\rm rad}$ are
universal to all orders in $ \alpha/\pi$ and the complete result reads
$\Delta g_{\rm rad} = (g_{\rm free} -2)\cdot [1+(Z\alpha)^2/6] +
\order{{\alpha\over \pi}\cdot (Z\alpha)^4}$.  The theoretical
uncertainty in the prediction for the experimentally interesting
carbon ion is decreased by a factor of about 3.
\end{abstract}


The interaction of an electron with  an
external magnetic field {\boldmath$B$} is described by the potential
\ba
V = -\vmu \cdot \mbox{\boldmath$B$}. 
\label{eq:pot}
\ea
The electron magnetic moment $\vmu$ is
\ba
\vmu = g_{\rm e}\, {e\over 2m} \mbox{\boldmath$s$}.
\label{eq:ge}
\ea
where $m$ and $\mbox{\boldmath$s$} = \vsig/2$ denote the mass and spin of 
the
electron and $g_{\rm e}$ is the so-called gyromagnetic or Land\'e factor.  We
adopt the convention that $e=-|e|$.

For a free electron, $g_{\rm e}$ is known with very high precision.  If an
electron is bound in a ground state of a hydrogen-like ion, $g_{\rm e}$
becomes a function of the nuclear charge $Z$ and its measurements
provide a sensitive test of the bound-state theory based on the
Quantum Electrodynamics (QED).  With a novel spectroscopic method
precise experiments can be carried out with
hydrogen-like ions in a wide range of nuclear charges $Z$
\cite{hermanspahn96,Quint95,Werth95}.   A unique
feature of those measurements is that results obtained with different
values of $Z$ may be used to rigorously test various bound-state
effects \cite{Karsh00}.

To fully exploit these experimental results, the QED prediction for
$g_{\rm e}\,(Z)$ must be known with comparable precision.  
At the present level of experimental uncertainty, accounting for the
QED interactions (including leading effects of the nuclear recoil) is
sufficient; other nuclear effects and weak interactions can be
neglected.  
The theoretical
prediction can be cast in the following form \cite{Mohr99}
\ba
g_{\rm e}\,(Z) = g_{\rm D} + \Delta g_{\rm rec} + \Delta g_{\rm rad}.
\ea
 
The first term corresponds to the lowest order expansion in
$\alpha/\pi$ and has been calculated to all orders in $Z\alpha$
\cite{Breit28},
\ba
g_{\rm D} = {2\over 3}\left[ 1 + 2\sqrt{ 1-(Z\alpha)^2 }\right].
\label{eq:Breit}
\ea

$\Delta g_{\rm rec}$ denotes the recoil corrections
\cite{grotch70recoil}, $\Delta g_{\rm rec} = 
\order{(Z\alpha)^2 {m\over m_N}}$, where $m_N$ is the nucleus mass.
Further references to the studies of those effects can be found in
\cite{Mohr99}. 

The main focus of the present paper are the radiative corrections.
They can be presented as an expansion in two parameters, $Z\alpha$ and
$\alpha/\pi$,
\ba
{ \Delta g_{\rm rad} \over 2}  =  
   C_{\rm e}^{(2)} (Z\alpha) \left( {\alpha\over \pi}\right)
 + C_{\rm e}^{(4)} (Z\alpha) \left( {\alpha\over \pi}\right)^2 
 + \ldots
\label{eq:series}
\ea
Powers of $\alpha/\pi$ correspond to electron--electron interactions,
while $Z\alpha$ governs binding effects due to electron interactions
with the nucleus.  The binding effects are relatively more important,
being enhanced by the nuclear charge and not suppressed by $1/\pi$,
peculiar to the radiative corrections.

The first coefficient function in (\ref{eq:series}), $C_{\rm e}^{(2)}
(Z\alpha)$, has been computed numerically to all orders in  $Z\alpha$
\cite{blundell97,Persson97}.  
Its first two terms in the $Z\alpha$ expansion are
also known analytically \cite{sch48,grotch70}
\ba
 C_{\rm e}^{(2)} (Z\alpha) = {1\over 2}\left[1 + {1\over 6}(Z\alpha)^2 +
 \order{(Z\alpha)^4}\right].
\label{eq:Grotch}
\ea
The main theoretical uncertainty for $g_{\rm e}$ in light ions is, at
present, connected with the unknown coefficient $C'$ in
the next coefficient function,
\ba
 C_{\rm e}^{(4)} (Z\alpha) &=&  C_{\rm e}^{(4)} (0)\left[1 + C'\cdot (Z\alpha)^2 +
 \order{(Z\alpha)^4}\right], 
\nonumber \\
 C_{\rm e}^{(4)} (0) &=& -0.328\,478\,444\,00\ldots \qquad
 \cite{som57,pet57a,Mohr99}.
\label{eq:cprime}
\ea

At present, the most accurate experimental value of the bound electron
gyromagnetic factor has been obtained
\cite{hermanspahn00,Haeffner} 
with a hydrogen-like carbon ion $^{12}$C$^{5+}$ ($Z=6$),
\ba
g_{\rm e}(Z=6;{\rm exp}) = 2.001\,041\,596(5).
\label{eq:exp}
\ea
  The theoretical prediction is \cite{Beier}
\ba
g_{\rm e}(Z=6;{\rm theory}) = 2.001\, 041\, 591(7)
\label{eq:theoryC}
\ea
where 70\% of the error is caused by the unknown coefficient $C'$ of
the $\left({\alpha\over\pi}\right)^2 (Z\alpha)^2$ effects in
(\ref{eq:cprime}) (for carbon, higher powers of $Z\alpha$ are assumed
to be negligible).

The purpose of this paper is to demonstrate that $C'={1/ 6}$, in
analogy to the corresponding coefficient in the lower order in
$\alpha/\pi$.  In fact, we will see that the coefficient of
$(Z\alpha)^2$ is the same in all coefficient functions
$C^{(2n)}_e(Z\alpha)$, so that the theoretical prediction for 
$\Delta g_{\rm rad}$ accurate up to $(Z\alpha)^2$ and exact in 
$\alpha/\pi$ reads:
\be
\Delta g_{\rm rad} = 
 (g_{\rm free} -2)\cdot
\left[1+ {(Z\alpha)^2\over6}\right],
\label{eq11}
\ee
where $g_{\rm free}$ is the gyromagnetic factor of a free electron,
presently known to $\order{(\alpha/\pi)^4}$ \cite{Hughes:1999fp}.
With this
result, the theoretical uncertainty in (\ref{eq:theoryC}) is reduced
from $7 \cdot 10^{-9}$ to about $2\cdot 10^{-9}$. 

To prove Eq.~(\ref{eq11}), we begin with a derivation of the
$\order{(Z\alpha)^2}$ term in the Breit correction (\ref{eq:Breit}),
working in full QED.  We will try to interpret the result in the
language of effective potentials whose average values give the
required correction.  In the next step we will construct from those
operators an effective Hamiltonian with which we will be able to
evaluate $(Z\alpha)^2$ corrections to higher orders in $\alpha/\pi$.

\begin{figure}[htb]
\hspace*{-1mm}
\begin{minipage}{6.cm}
\[
\mbox{
\begin{tabular}{cc}
\psfig{figure=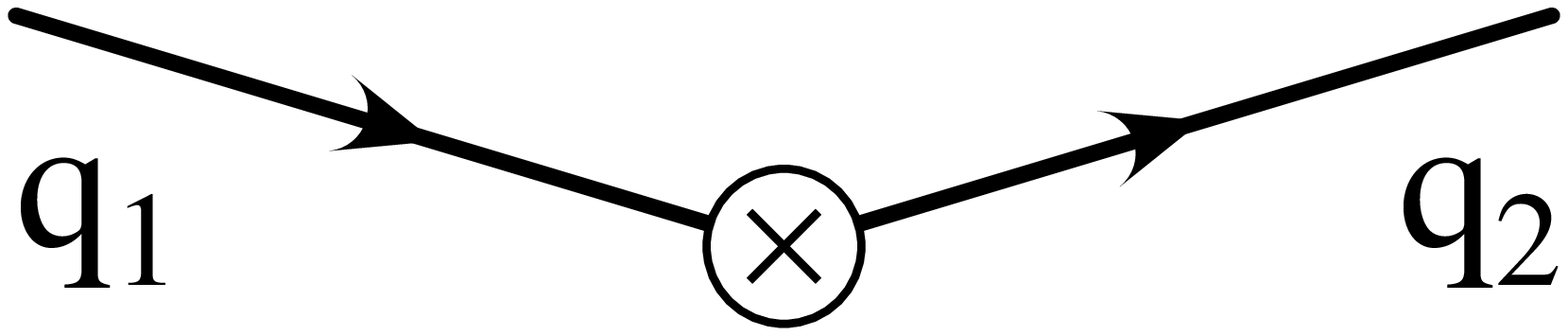,width=35mm}
&\hspace*{5mm}
\psfig{figure=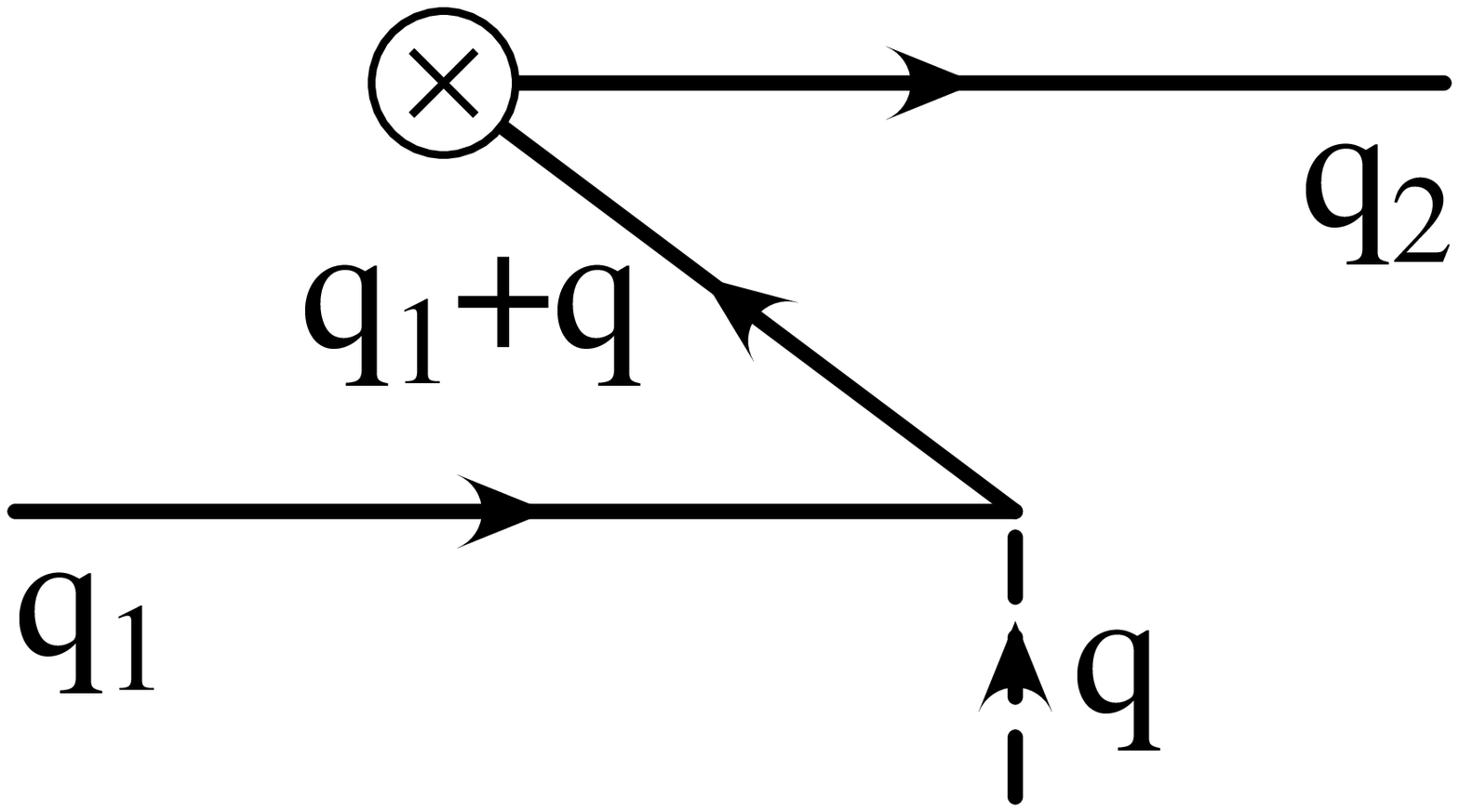,width=35mm}
\\[1mm]
 (a) & (b)
\\[1mm]
\end{tabular}
}
\]
\end{minipage}
\caption{\sf Tree-level contributions to the factor $g$ of a bound
electron.  The cross denotes the magnetic field insertion, and the
dashed line is the interaction with the Coulomb field.  The diagram
(b) has a counterpart with the electric and magnetic fields
interchanged.}
\label{fig:Lande}
\end{figure}

There are two contributions which have to be considered, shown in
Fig.~\ref{fig:Lande}.  The velocity of the electron in the ion is of
the order of $Z\alpha$; in order to compute corrections $(Z\alpha)^2$,
it is sufficient to expand the matrix elements to second order in
electron momentum, relative to the leading term.

The diagram \ref{fig:Lande}(a) describes the scattering of an electron
on the magnetic field $\boldB$.  
Expanding this matrix element with respect to electron's velocity we
arrive at the following effective potential:
\ba
\Delta V_a &&= -\frac {e}{2m} \left \{
\vsig \cdot \boldB \left ( 1 - \frac {\p^2}{2m^2} \right  )
\right. \nonumber \\
&& \left. 
+ \frac {i}{4m} \left [ H+\frac{Z\alpha}{r}, 
\vsig \cdot \left ( \A \times \p - \p \times \A \right ) \right ] 
\right \},
\ea
where $H = \p^2/2m - Z\alpha/r$ is the non-relativistic Hamiltonian.
Since $ \langle \Psi | [H,U]| \Psi \rangle = 0$ for any operator $U$, 
we find the following expression for $\Delta V_a$
\ba
\Delta V_a &&=
-\frac {e}{2m} \left \{
\vsig \cdot \boldB \left ( 1 - \frac {\p^2}{2m^2} \right  )
\right. \nonumber \\
&& \left. 
- \frac {e}{4m}  
\vsig \cdot \left ( \A \times \boldE - \boldE \times \A \right ) 
\right \}.
\label{eq:Va}
\ea

The Z-diagram in Fig.~\ref{fig:Lande}(b) describes a transition of the
electron into the negative energy sea after interacting with either
magnetic or electric field.  Since the energy
of the intermediate state is of the order of the electron mass, this
is a short distance process and it can be described by a local
operator,
\be
\Delta V_b =-\frac {e^2}{4m^2}
\vsig \cdot \left ( \A \times \boldE - \boldE \times \A \right ). 
\label{eq:Vb}
\ee

The sum of contributions (\ref{eq:Va}) and (\ref{eq:Vb}) reads
\ba
\Delta V &&=
-\frac {e}{2m} \left \{
\vsig \cdot \boldB \left ( 1 - \frac {\p^2}{2m^2} \right  )
\right. \nonumber \\
&& \left. 
+ \frac {e}{4m}  
\vsig \cdot \left ( \A \times \boldE - \boldE \times \A \right ) 
\right \},
\label{eq15}
\ea
and, after being averaged over  the $1S$ state,  gives  
the leading binding correction to the $g_{\rm e}$-factor:
\be
g_{\rm D} \simeq 2 -\frac {2(Z\alpha)^2}{3},
\label{gD}
\ee
which agrees with first two terms of expansion of Eq.~(\ref{eq:Breit}).

We remark that there could be another source of 
$(Z\alpha)^2$ corrections
induced by Breit potential $V_{\rm Breit}$,
$$
  \langle 1S |\vsig \cdot \boldB  
\sum \limits_{n \ne 1} 
\frac {|n \rangle  \langle n |}{E-E_n} V_{\rm Breit}| 1S \rangle.
$$
However, this expression vanishes for the constant magnetic field
$\boldB$, and therefore the result in (\ref{gD}) is complete.

Although Eq.~(\ref{eq15}) leads to a correct result, its second term
depends on the electromagnetic potential $\A$ and therefore is not
gauge invariant.  How is the gauge invariance restored? The answer is
that there should be an additional $\A$-independent contribution to
the potential, so that the second term in Eq.~(\ref{eq15}) is a part of
an explicitly gauge invariant expression
\be
\frac {e}{8m^2}
 \vsig\cdot 
\left( 
 \PP \times {\boldE} 
 -  \boldE \times \PP 
\right ),
\ee
where 
$\PP = \p - e\A$ 
is the canonical momentum of the Coulomb Hamiltonian. 
Although this operator
depends on both $\boldE$ and $\A$, its coefficient 
can be determined more easily by switching off the magnetic field and 
considering the scattering of an electron on an electric field alone.

The above considerations suggest the form of the
general effective Hamiltonian describing the interaction of an
electron with a magnetic field {\boldmath$B$}, in the presence of an
electric field {\boldmath$E$},
\ba
\Delta H &=& 
-c_0 {e\over 2m}  \vsig \cdot \boldB
+
 c_1 {e\over 8 m^2} {\vsig \cdot  \left ( \boldP \times \boldE 
- \boldE \times \boldP \right )}
\nonumber \\
&&+ c_2 \frac {e}{4m^3}\, \boldp^2\; \vsig \cdot \boldB
 + c_3 \frac {e}{4m^3}\,  \boldp \cdot \boldB \; \vsig\cdot \boldp .
\label{heff}
\ea
The coefficients $c_i$ can be found by the standard procedure of
matching the amplitudes obtained within the effective theory with
those in the full QED \cite{Caswell:1986ui}.
To this end, we consider the on-shell elastic
scattering of an electron on the electric field (for $c_1$) or on the
magnetic field (for $c_{0,2,3}$).  The interaction of an electron
with the electromagnetic field is described by two form factors (in the
absence of parity violation),
\ba
\lefteqn{
\langle e(p') | J^{\rm em}_\mu | e(p) \rangle }
\nonumber \\ &&
  = e \, \bar u_e(p') \left(
 F_D(q^2) \gamma^\mu
 + {i\over 2m} F_P(q^2) \, \sigma^{\mu\nu} q_\nu
\right)
u_e(p).
\label{eq:ff}
\ea
Here $q=p'-p$. 
For the determination of $c_i$ we can treat the external fields as
constant, and need the form factors only at $q^2=0$.  In this case the
Dirac form factor is $F_D(0)=1$ and the Pauli form factor gives the
anomalous magnetic moment $F_P(0) = \alpha/2\pi +
C^{(4)}_e(0)\,(\alpha/\pi)^2 + \ldots$.  

The form of interaction (\ref{eq:ff}) determines the scattering
amplitudes in external magnetic and electric fields, and we easily
find the coefficients of the relevant operators in (\ref{heff}),
\ba
c_0 &=& F_D + F_P,
\qquad
c_1 = F_D+2F_P,
\nonumber \\
c_2 &=& F_D, \qquad  c_3 = F_P.
\ea

To find the interaction energy of a bound electron in an external
magnetic field, we compute the expectation value of
 the Hamiltonian  (\ref{heff}) in the $1S$ state and find
\ba
V &=& -{e\over 2m} (F_D+F_P)  \vsig \cdot \boldB
\nonumber \\ &&
 - {e\over 2m} (Z\alpha)^2 \left( {F_P - 2F_D\over 6} \right) \vsig
 \cdot \boldB 
 + \order{(Z\alpha)^4}.
\ea
Comparing this result with (\ref{eq:pot}) and (\ref{eq:ge}) we find
(neglecting nuclear recoil and effects of $\order{(Z\alpha)^4}$)
\ba
{g_{\rm e}(Z)\over 2} = F_D+F_P + (Z\alpha)^2 {F_P-2F_D\over 6}.
\ea
This result is valid to all orders in the ``radiative'' expansion
parameter $\alpha/\pi$.  We can re-write it as
\ba
{g_{\rm e}(Z)\over 2} &=& \left( 1- {(Z\alpha)^2\over 3}\right)
+F_P \left( 1+ {(Z\alpha)^2\over 6}\right)
\nonumber \\
&=& {g_{\rm D}\over 2} +
\left[ {\alpha\over 2\pi}
 + C_{\rm e}^{(4)}(0)\,\left( {\alpha\over \pi} \right)^2 + \ldots
\right]
\nonumber \\ &&
\times \left( 1 +  {(Z\alpha)^2\over 6}\right) + \order{(Z\alpha)^4}.
\ea
Comparing this with eqs.~(\ref{eq:Breit}, \ref{eq:Grotch}), 
we see that we have correctly reproduced the known 0- and 1-loop
results.  We have also found, that the bound-state correction factor
$C'={1/ 6}$ is universal in all orders in $\alpha/\pi$.  

\subsection*{Summary}
We have demonstrated a relation between the gyromagnetic factors of
free and bound electrons, valid to the lowest order in $(Z\alpha)^2$
and to all orders in $\alpha/\pi$.  The main reason for this somewhat
unexpected relation is that only low-dimensional effective operators 
contribute
to order $(Z\alpha)^2$.  

We do not anticipate a similar relation involving only known
form factors $F_D$ and $F_P$ to hold for the higher order binding
effects.  For example, in $\order{(Z\alpha)^4}$, the operator $\vsig
\cdot \boldB\, (\boldE\cdot\boldp + \boldp\cdot\boldE)$ might contribute.  In
general, the coefficients of such operators in the effective
Hamiltonian are new functions of $\alpha/\pi$, independent of
$F_{D,P}$.

In the previous theoretical prediction for the bound electron $g_{\rm e}$
\cite{Beier}, the main source of uncertainty was the unknown two-loop
binding effect, which has been estimated as $3\alpha/\pi$ times the 
one-loop binding effect.  For $Z=6$ it is $3(\alpha/\pi)^2\cdot
(Z\alpha)^2/6 \simeq 5\cdot 10^{-9}$.  Together with the error in the
nuclear recoil, the total theoretical uncertainty was estimated as $7
\cdot 10^{-9}$.  The result of the present
paper, which gives the explicit two-loop binding effect, shifts the
central value of the theoretical prediction, eq.~(\ref{eq:theoryC}),
by $-1.13\cdot 10^{-9}$, and reduces its uncertainty by a factor of
about 3.  The remaining uncertainty is dominated by the errors of the
recoil correction $\Delta g_{\rm rec}$ and of the numerical evaluation
of the binding effects in the order ${\alpha/\pi}$ (see \cite{Beier}
for a detailed discussion).  With this reduction in the theoretical
uncertainty, a very precise value of the electron mass can be
extracted from bound electron $g$-factor experiments \cite{werthpc}.

Finally, let us note that a confrontation of the theoretical
prediction (\ref{eq:theoryC}) with the experimental results
(\ref{eq:exp}) for $g_{\rm e}$ tests the bound-state QED at the level
of 1\%.  For comparison, measurements of the positronium hyperfine
splitting test the bound-state QED effects at the level of 0.3\%
\cite{Ph,PhK,Czarnecki:1998zv}.  If the experimental uncertainty in
the bound electron $g$-factor can be further reduced, its
measurements, combined with an independent electron mass
determination, will rank among the most stringent tests of the
relativistic bound-state theory.

{\it Note added.}  After completing this work, we learned about
Ref.~\cite{EG}, where $\order{(Z\alpha)^2}$ corrections were also
considered. Although our approach differs substantially from that of
Ref.~\cite{EG}, the final results agree.  Our paper should, therefore,
be viewed as an alternative way of deriving the binding
$\order{(Z\alpha)^2}$ corrections to the gyromagnetic factor of the
electron.  We believe, however, that the results of Ref.~\cite{EG}
have not become well known; for this reason our alternative derivation
and the emphasis on phenomenological consequences seems to be timely.

\subsection*{Acknowledgments}
We thank M. Eides, S. Karshenboim, P. Mohr, K. Pachucki,
S. Salomonson, J. Sapirstein, and G. Werth for discussions and
comments on the manuscript.  A useful conversation with Stan Brodsky
is gratefully acknowledged.  We are grateful to the BNL High Energy
Theory Group for hospitality during the work on this problem.  We
would like to thank particularly Dr. Sally Dawson and Professor Vernon
Hughes for providing support, which made our visit at BNL possible.
This work was supported in part by DOE under grants number
DE-AC02-98CH10886 and DE-AC03-76SF00515, and by the Russian Foundation
for Basic Research grant 00-02-17646.


\end{document}